\documentstyle[12pt,fleqn]{article}
\begin{document}

\begin{center}
{\large \bf Can Temperature Be Quantized?}

\vskip 0.1in
Li-Xin Li

{\it Department of Physics, Beijing Normal University, Beijing 100875,
China} 
\end{center}

\vskip 0.3in
\begin{center}
\begin{minipage}{6.0in}
{\bf Abstract:} It is shown that the temperature may be quantized in
some spacetime due to the periodically topological structure of the 
Euclidean section. The quanta of the temperature is the Hawking-Unruh 
temperature.
\end{minipage}
\end{center}

\section*{} 
~~
In the Euclidean quantum theory physical functions in the Lorentzian 
section of a complexified spacetime are assumed to be analytically 
continued from some functions in the Euclidean section [1, 2]. The 
non-trivial topological structure of the Euclidean section may cause 
some interesting effects in the Lorentzian section. For example, 
people have found that in a Lorentzian section whose corresponding 
Euclidean section is periodic in the Euclidean time $\tau=it$  there 
is the Hawking-Unruh effect (such as in the black hole spacetime and 
the de Sitter spacetime). In such a spacetime (Here we call it the 
Lorentzian Hawking-Unruh type spacetime (L-HU-spacetime), and call 
the corresponding Euclidean section the Euclidean Hawking-Unruh type 
spacetime (E-HU-spacetime)), observers whose worldlines are the
integral curves of $(\partial/\partial t)^a$, feel that they are in a 
thermal bath with the temperature $T_0=1/\beta_0=\kappa/2\pi$, where 
$\beta_0$ is the period of the Euclidean time and $\kappa$ is the 
surface gravity of the event horizon [3 - 5]. In this letter I show 
that in a L-HU-spacetime the temperature may be quantized with the 
quanta $T_0$ which is the lowest possible temperature for the thermal 
equilibrium.

It is well known that the Euclidean thermal Green function
$G_T(\tau,\vec{x};\tau^\prime, \vec{x}
^\prime)$ is a periodic (for bosons) or 
anti-periodic (for fermions) function of $\tau$ (and $\tau^\prime$) 
with the period $\beta=1/T$ where $T$ is the temperature [6, 7]. This 
holds for systems with vanishing chemical potential in a stationary 
spacetime or a conformally stationary spacetime [7]. It is easy to 
show that $\beta$ is the {\it fundamental period} of such functions. 
For example, for the scalar field, $G_T(\tau,\vec{x};
\tau^\prime, \vec{x}^\prime)=i{\rm tr}[e^{-\beta H}{\cal
T} (\phi(\tau,\vec{x})\phi(\tau^\prime,\vec{
x}^\prime))] /{\rm tr}(e^{-\beta H})$ where ${\cal T}$ denotes the 
Wick time-ordering. $G_T(\tau,\vec{x};\tau^\prime,\vec{x}
^\prime)=G_T(\tau+\alpha,\vec{x};\tau^\prime, \vec{x}
^\prime)$ ($0<\alpha\leq\beta$) leads to
$\sum_{mn}(e^{(\alpha-\beta)E_m-\alpha E_n}-e^{-\beta
E_n})\vert\langle m\vert\phi(\tau,\vec{x})\vert
n\rangle\vert^2=0 $ ($H\vert m\rangle=E_m\vert m\rangle$ and $\langle
m\vert n\rangle=\delta_{mn}$) in the limit
$\tau^\prime\rightarrow\tau$ and $\vec{
x}^\prime\rightarrow \vec{x}$, which immediately results
in $\alpha=\beta$.

In an E-HU-section with the period $\beta_0$ in the Euclidean time 
$\tau$, every thermal Green function for bosons should be periodic 
with respect to the translation $\tau\rightarrow\tau+\beta_0$ [4, 7]. 
While it is also a periodic function of $\tau$ with the {\it fundamental 
period} $\beta$, we must have $\beta_0=n\beta$ ($n = 1, 2, ...$)
because every doubly-periodic function reduces to a singly-periodic 
function when the ratio of the two periods is real [8]. Such a
conclusion also holds for fermions since an anti-periodic function
with the period $\beta$ is also a period function with the period
$2\beta$ and every spinor thermal Green function for fermions should 
be anti-periodic with respect to the translation
$\tau\rightarrow\tau+\beta_0$ [7]. Therefore the allowed temperature 
in the Lorentzian section is
\begin{eqnarray}
T_n=nT_0,~~~n=1,2,...\nonumber
\end{eqnarray}
with $T_0=1/\beta_0=\kappa/2\pi$, which means that in the
L-HU-spacetime, the temperature should be quantized with quanta 
$T_0$, and $T_0$ is the lowest possible temperature for the thermal
equilibrium.

\end{document}